\documentclass{aa}
\usepackage{graphicx}
\usepackage{natbib}
\bibpunct{(}{)}{;}{a}{}{,}

\begin{document}

\title{Null-induced mode changes in PSR~B0809+74}

\author{
A.G.J. van Leeuwen     \inst{1}
\and
M.L.A. Kouwenhoven     \inst{1}
\and
R. Ramachandran        \inst{2,3}
\and
J. M. Rankin           \inst{2,4}
\and 
B.W.~Stappers          \inst{3}
}
\institute{Astronomical Institute, Utrecht University, PO Box 80000,
  3508 TA Utrecht, The Netherlands
  \and
  Astronomical Institute `Anton Pannekoek',
  Kruislaan 403, 1098 SJ Amsterdam, The Netherlands  
  \and
  Stichting ASTRON, PO Box 2, 7990 AA Dwingeloo, The Netherlands
  \and
  Physics Department, University of Vermont, Burlington, VT 05405,
  USA
}
\offprints{a.g.j.vanleeuwen@astro.uu.nl}
\date{Received / Accepted}

\abstract{We have found that there are two distinct emission modes in
PSR~B0809+74. Beside its normal and most common mode, the pulsar emits
in a significantly different quasi-stable mode after most or possibly
all nulls, occasionally for over 100 pulses. In this mode the pulsar
is brighter, the subpulse separation is less, the subpulses drift
more slowly and the pulse window is shifted towards earlier longitudes.
\newline We can now account for several previously unexplained phenomena
associated with the nulling-drifting interaction: the unexpected
brightness of the first active pulse and the low post-null
driftrate. We put forward a new interpretation of the
subpulse-position jump over the null, indicating that the speedup
time scale of the post-null drifting is much shorter than previously
thought. The speedup time scale we find is no longer discrepant with
the time scales found for the subpulse-drift slowdown and the emission
decay around the null.
}

\maketitle

\section{Introduction}

Lately, it has been quiet around bright PSR~B0809+74, once the
canonical example of regular subpulse drifting and the centre of
lively discussions.

\begin{figure*}[tb]
	\centering
	\includegraphics[]{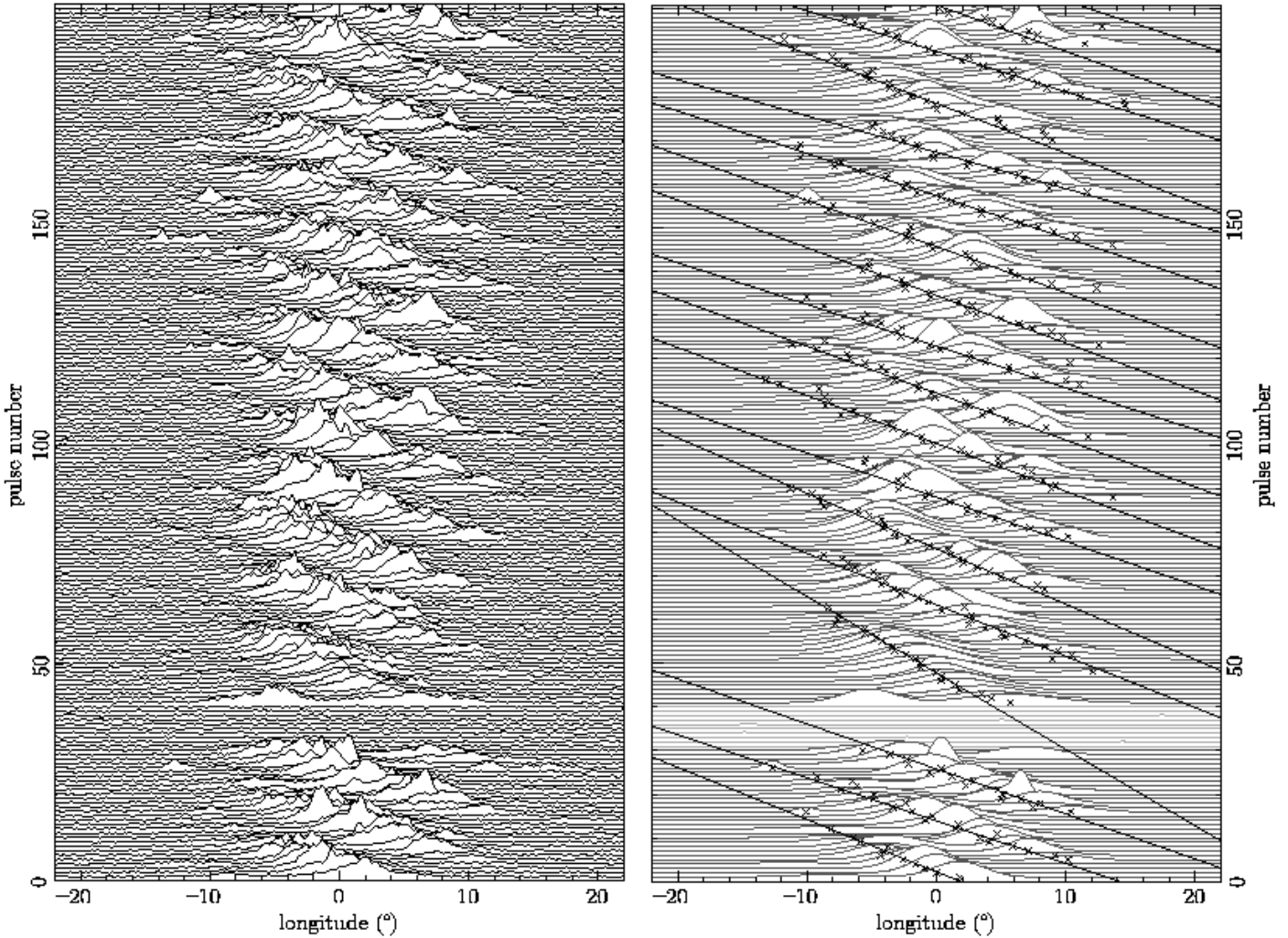}
	\caption{Example of the fitting of the subpulses.  Left panel:
		stacked pulses in the on-pulse region.  Right panel:
		after finding nulls (lighter colour), and fitting
		subpulses (crosses at the base of the fitted
		Gaussians), we fit the driftbands (black lines). In
		this and all other figures, 0$^{0}$ longitude is
		defined as the centre of the best Gaussian fit to
		the average profile.}
		\label{img:fit.stack}
	\end{figure*}

In the thirty years since their discovery by \citet{dc68}, drifting
subpulses feature in many discussions about the nature of pulsars and
the pulsar's emission mechanism. The subpulse-drifting phenomenon in
itself is simple: when comparing adjacent individual pulses, the
subpulses that comprise them are seen to shift regularly through the
average pulse window. The left panel of Fig.  \ref{img:fit.stack}, for
example, shows a recent observation in which these subpulses form
their driftbands.

In 1968, the interpretation of Drake \& Craft that the subpulse
drifting is the pulsation of the neutron star, is ground zero for much
debate.  Over the following years, the ever increasing quality and
volume of observations of subpulses in different pulsars continue to
limit possible models and require their refinement. It is in this
process that observations of 0809+74 take the lead.

\citet{avs69} and \citet{vs70} are the first to detect the driftbands of
0809+74. \citet{col70a} then notes that the driftrate of the
subpulses occasionally changes, without being able yet to identify the
nulls as the trigger. One year later, \citet{th71} do find that this
occasional cessation of the pulsar's emission precedes a changing
driftrate, lasting a few driftbands.

In his paper solely devoted to the drifting in 0809+74,
\citet{pag73} finds that the subpulse separation changes through the
profile. He also introduces the `subpulse phase', the
difference between the actual position of the subpulse and its
predicted position. While this makes it easier to plot long series of
subpulse positions horizontally, it also introduces the notion that
the subpulse position dramatically jumps in phase 
over a null. From the data however, it is clear that the subpulse position
over the null does not change much at all -- on the contrary, it is the
preservation of phase that is the astonishing phenomenon.

\citet{urwe78} are the first to note this. They find that the position
of the subpulses is identical before and after the null and conclude
that, even though there is no emission from the pulsar, the emitting
structures survive the null. Five years later, \citet{la83} use a
method devised by \citet{rit76} to find the nulls of 0809+74 and
investigate the time scales associated with the change in drift
pattern around a null. The first time scales they look into are those
of the decay and rise of emission around a null. A finite decay and
rise time would cause the pulses neighbouring a null to be less bright
than the average. A relative dimness of the last pulse before the null
is found indeed, pointing at a decay time shorter than 5\% of the
pulse period. Contrary to the expectations, however, the first pulses
after the nulls are found to outshine the average ones.

The next time scales considered are those of the slowdown and speedup
of the subpulse drift around the null. Lyne \& Ashworth notice that,
contrary to the findings of Unwin et al., there is a some change in
subpulse positions over the null. They suggest that this change
originates in subpulse drift that is caused by a gradual speedup of
the drifting during the null. Being several tens of pulse periods,
this speedup time scale is much larger than the time scales found for
the emission decay and rise, and the subpulse drift slowdown.

In 1984, simultaneous observations at 102 and 1412 MHz by
\nocite{dls+84} Davies et al.~confirm the curvature of the driftbands
found by \citet{pag73} and show that the subpulse width varies across
the pulse profile.

Although work on 0809+74 quiets down after this intense period,
significant observational progress for the understanding of subpulse
drifting in general is made on 0943+10. \citet{dr99, dr01} are
able to detect periodicities in the subpulse strengths that lead to a
determination of the recurrence time of individual subpulses.

On the theoretical front, the initial hypothesis that the subpulse
drifting originates in pulsations of the neutron star \citep{dc68} is
soon abandoned. The subpulses are now thought to be formed in the
magnetosphere of the pulsar, and the topic of debate changes to the
exact location of their formation. The observed periodicity
is first thought to be caused by quasi-periodic fluctuations in the plasma
near the co-rotation radius of the pulsar. The polarisation and
beaming of the subpulses, however, later tie the emission mechanism to
the surface of the pulsar.

In 1975, Ruderman \& Sutherland \nocite{rs75} build on the
\citet{gj69} and \citet{stu71} models to propose a theory that
incorporates the drifting subpulses and the nulling. They suggest that
the pulsar emission is formed in discrete locations around the
magnetic pole. These locations would rotate around the pole like a
carousel.  In one of many revisions of the Ruderman \& Sutherland
model, \citet{fr82} propose a model that explains the survival of the
subbeam structure over the null.  The work of \citet{dr99, dr01}
visualises these models by mapping the observed subpulses onto their
original positions on the carousel.

In this paper we investigate the behaviour of the pulsar emission around
nulls. We will use observations of two extraordinary long sequences of
post-null behaviour to explain the features of general subpulse
drifting around nulls in a new framework.

\section{Observations}
\subsection{Data}

Over a period of about 18 months we have observed 0809+74 with
PuMa, the Pulsar Machine, at the Westerbork Synthesis Radio Telescope
(WSRT) in the Netherlands.  Most observations lasted several times the
scintillation time scale of this pulsar. We have only used the data in
which the pulsar was bright, amounting to a total of 13 hours. With a
pulse period of 1.29 seconds, this comes down to $3.6 \times 10^4$
pulses.

The collecting area of WSRT and the low system noise made it
possible for PuMa to record the data with a high timing resolution and
a high signal-to-noise ratio (SNR). The WSRT consists of fourteen
25-meter dishes arranged in east-west direction. For our observations,
signals from all the fourteen dishes were added after compensating for
the relative geometrical delay between them, so that the array could
be used like a single dish with an equivalent diameter of 94
meters. The pulsar was observed with PuMa \citep{kou00, vou01, vkh+02}
at centre frequencies between 328 MHz and 382 MHz, with bandwidths of
10 MHz. Each 10-MHz band was split into 64 frequency channels, and the
data was recorded, after being digitised to represent each sample by
four bits (sixteen levels). After inspection for possible electrical
interference, we corrected for the interstellar dispersion during our
off-line analysis.

\subsection{Reduction}
\subsubsection{Finding nulls}

	\begin{figure}[tb]
	\includegraphics[]{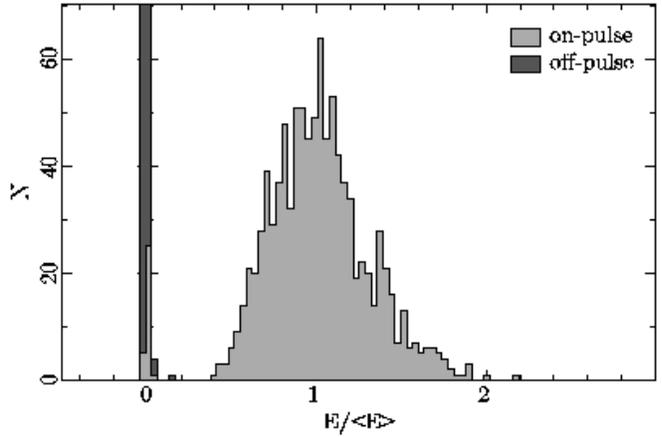}
	\caption{Histogram of the energy in 1000 consecutive pulses,
	each scaled with the average energy of the 100 pulses that
	surround it. In the foreground (light gray) we show the
	distribution of the energies found in the on-pulse window. In
	the background (dark gray) the energies found in an equally
	large section of the off-pulse window are shown, peaking far
	off the scale around N=900.}
	\label{img:null.hist}
	\end{figure}	

Normally, the energy of individual pulses varies from about 0.5 to 2
times the average. When the pulsar is in the null state however, which
it is for about 1.4\% of the time, it is does not emit. In this case,
very little energy is found where the pulse is expected. We have used
this difference in on-pulse energies to identify the nulls. To do
this, we compared the on-pulse energies with the energy found in
off-pulse region, where no emission is expected.

To calculate the on-pulse and off-pulse energies, we split each pulse
period $P_1$ (360$^{o}$ in longitude) into three sections. The first
section, $0.15 P_1$ wide, was centred around the peak of the average
profile. The second section, equally wide, consisted of a part of the
off-pulse region. A third section, $0.5 P_1$ wide, contained a
different, independent part of the off-pulse data and was used to
estimate the noise and baseline of the signal. This baseline was
subtracted from both the on-pulse and off-pulse data.

To calculate the energy in each pulse, we determined the central range
in longitude that on average contained 90\% of the power the pulsar
emitted in the on-pulse section.  For this region we summed the
amplitudes of the signal to get the energy content. We did the same
for an equally large piece of the off-pulse section. To correct for
long term variations in the pulse energy due to scintillation, we
subsequently scaled both energies with the average energy of the 100
surrounding pulses.

We defined that, in the null state, there is no observable radiation
from the pulsar. This means that noise is all one observes at the time
a new pulse is expected. The energy distribution of the null-state
is then identical to that of the off-pulse region (high
distribution in the background of Fig. \ref{img:null.hist}). For our
data this meant that all pulses with energies less than the highest
energy found in the off-pulse distribution were nulls (small peak
around zero energy in Fig. \ref{img:null.hist}). Furthermore, the
off-pulse distribution gives a better estimate of the expected spread
in null energies than the null distribution itself, due to the larger
number of pulses included.

This method, originally devised by \citet{rit76}, was first used on
0809+74 by \citet{la83}. In their case, there was considerable
overlap between the on-pulse and off-pulse distributions, hindering
their identification of the nulls. We wanted to be certain that the
set of nulls we found was genuine and complete; therefore we only used
the 60\% of our observations in which the on-pulse and off-pulse
distribution were separated by more than $0.1 E/\!\!<\!\!E\!\!>$.

With a 3-sigma point at 15 mJy, the low noise makes it possible
to fully distinguish between pulses in the normal state on the one
side, and those in the null state on the other, giving unprecedented
insight into the nature of the null state.

\subsubsection{Fitting subpulses}
The major step in our data reduction was to extract the four basic
elements of the individual pulses: the number of subpulses and their
respective positions, widths and heights. By doing so we greatly
increased the speed of subsequent operations, but at the same time
kept a handle on the physics by retaining the parameters most
important for visualising the data.

The high SNR of our observations showed a wealth of microstructure in
the individual subpulses, which on average were Gaussian in shape.
Plainly fitting Gaussians to the intricate, many featured subpulses
did not immediately result in locating them reliably. Due to the
changing amplitude of the signal throughout the window, the
microstructure of the brightest subpulses was often more important to
the goodness-of-fit than complete, weaker subpulses near the edge of
the pulse window. In order to still find these weak subpulses, we
devised a two-step approach of first finding the strong subpulses and
then using their driftpath to suggest the position of weaker
subpulses. This approach worked very well.

For our fitting, we assumed the subpulses to be Gaussian in shape, non
overlapping, and to have a full width at half maximum (FWHM) less than
15$^{o}$. We used a Levenberg-Marquadt method \citep{ptvf92} with
multiple starting configurations to produce goodness-of-fit values for
different numbers of subpulses. By comparing these $\chi^2$ values, we
located the subpulses that had high significance levels.

Upon finding these normal to strong subpulses, we set out to detect
the weaker ones over the noise. For this purpose we composed
driftbands out of the individual subpulses already identified. On a
one-by-one basis, a subpulse was either added to a path that had
predicted its position to within $P_2/4$ ($P_2$ being the average
separation between two subpulses within a single pulse, about $11^{o}$
longitude), or taken to be the start of a new path. Paths ended if no
subpulse fitted the path for more than 5 pulses, or upon reaching a
null. On the first run, only paths longer than ten pulses were allowed
to survive, so as to eliminate the interference of short run-away
paths. This allowed a steady pattern of long paths to grow,
incorporating about 90\% of the subpulses found. On the second run,
paths longer than 5 pulses were allowed to form: these shorter paths
occur only around nulls, where the normal, long paths are interrupted.

The drifting pattern thus formed predicted the locations of the weaker
subpulses. Around each predicted position, we isolated a section of
the data. This section was taken as large as possible without
interfering with other, previously fitted subpulses. We checked the
significance of fitting a single Gaussian to indicate the presence of
a subpulse.

In the same way as described above, the final driftpattern was
then identified using the extended set of subpulses.

To check the method, we compared the observed average profile with the
one recreated from the Gaussians fits to the subpulses. From their
match, we concluded that the fitting procedure was effective.

\section{Results}
\subsection{Driftband fitting}
\begin{figure}[tb]
	\includegraphics[]{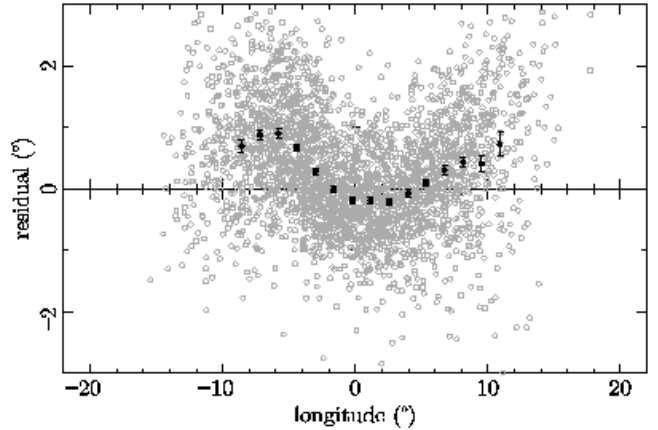}
	\caption{Horizontal residuals to straight line fits to the
	driftbands, individual and binned, for 2000 pulses.}
	\label{img:resid}
	\end{figure}

The driftbands are not straight, but show a systematic curvature
\citep{dls+84}. This is seen most strikingly when we plot the
residuals to straight line fits: Figure \ref{img:resid} shows the
longitudinal offset between the actual and predicted positions of the
subpulses. Near the edges of the profile the subpulses arrive later
than is to be expected based on a linear driftband, in the middle they
arrive sooner.

This probably explains the results of \citet{ps82}. After fitting straight
lines to all driftbands, they find that the driftrate of the last
driftband before the null is 20\% higher than that of a normal
driftband. As seen in the curvature of the driftbands, however, a
normal driftband consists of a fast drifting first half and a slow
drifting second half. If a driftband is cut by a null, the part before
the null consists only of the fast drifting first half, resulting in a
higher average driftrate over this shorter driftband.

To make our results independent of the longitude, the non-linearity of
the driftbands is taken into account in all the following driftrate
calculations. In these cases, the positions of individual subpulses are
corrected by subtracting the appropriate residual value.

\subsection{Nulling}
\subsubsection{Shortest nulls}

	\begin{figure}[tb]
	\includegraphics[]{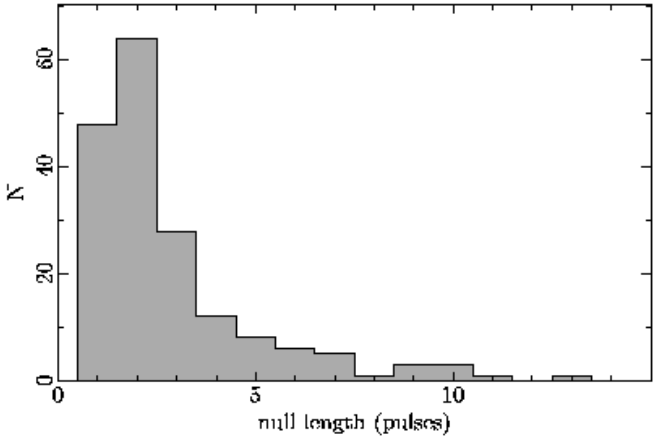}
	\caption{Length histogram for the 180 nulls.}
	\label{img:null.length}
	\end{figure}	

The main criterion used to decide whether or not to include certain
datasets was the complete separation of the on-pulse and off-pulse
energy distributions. Therefore, the set of null identified is genuine
and complete. This allows us to investigate the underlying statistics
of null occurrence and length, and estimate the influence of short
nulls.

The chance of finding 0809+74 in the null state is on average
about 1.4\%. If the null lengths were distributed in a Poissonian
fashion, we would find a preponderance of one-pulse nulls and very few
longer ones. However, in the null-length histogram
(Fig. \ref{img:null.length}) we find a peak at nulls of length 2, a
significant decrease towards shorter lengths and a considerable number
of long nulls, showing that the occurrence of nulls is not governed by
pure chance.

Comparing this histogram with the one previously found by
\citet{la83}, we see that we identify about twice as many one-pulse
nulls in the data. Still, many nulls shorter than one period must pass
unnoticed, as they occur when the pulsar faces away from us.

We know that nulls, especially long ones, have a distinct impact on
the drifting pattern in their vicinity. Although the impact of shorter
nulls is less, a large number of unnoticeable, short nulls ($< P_1$)
might seriously influence the drifting pattern we are trying to
understand.  

Using the null-length histogram we can estimate the
number of these short nulls. It peaks at two-pulse nulls, and the
distribution decreases towards shorter lengths.  Assuming that the
underlying distribution of null-lengths is continuous, this tendency
of decreasing occurrence towards shorter nulls implies that there is a
small number of nulls shorter than one pulse period. Extrapolating
the decrease leads to an estimate of about 15 unnoticeable short
nulls in the null-length interval from 0 to 0.5 pulses.

The low frequency of their occurrence (0.04\%) indicates the influence of
short nulls on the drifting pattern is negligible.

\subsubsection{Null versus burst length}
The next question we address involves the interval between adjacent
nulls (the so-called burst) and the duration of the nulls. Does
waiting longer for a null mean it will last longer, too?  We have
checked these relationships and have found that the lengths of
neighbouring nulls and bursts are independent.

\subsubsection{Position jump over nulls}

	\begin{figure}[b]
	\includegraphics[]{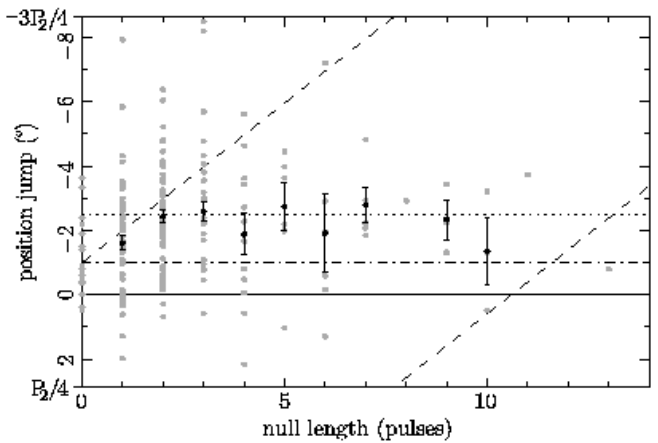}
	\caption{Jump in subpulse position against null length. The
	bottom of the plot falls at $P_2/4$, the top at $-3P_2/4$. The
	diagonal dashed line is the predicted subpulse path if the
	drifting were independent of the nulls. The horizontal
	dash-dotted line is the predicted path for a sudden and
	complete stop of drifting during the null. The horizontal
	dotted line is the average jump found for the nulls longer
	than 1 period.}
	\label{img:jump.length}
	\end{figure}	

If the nulling mechanism is independent of the position of the
subpulses, we expect that the distribution of subpulse positions is
the same for the normal pulses and the pulses that immediately precede
a null. We find no proof of differences in these distributions and
conclude that there is no preference for a null to start at a certain
subpulse position.

The positions of subpulses change over a null. We derived this shift
of the subpulse pattern for each null in our sample. As the positions
of the individual subpulses were already identified, this shift was
simply extracted. Each subpulse before the null matched a subpulse
after the null, if the latter fell within $-3P_2/4$ to $P_2/4$ of the
former. This range is symmetric around the average expected jump over
a null, so as to minimise the number of ambiguous cases. In the
following analysis, we used the average of all individual subpulse
shifts within one pulse.

Figure \ref{img:jump.length} shows this shift in the subpulse position
over a null, corrected for the non-linear behaviour of the
driftband. If the motion of the subpulses were independent of the
emission, the subpulses would continue to drift invisibly throughout
the null state, and reappear at a very different position. The
associated shift in positions would then be spread around the diagonal
dashed line in Fig. \ref{img:jump.length}.

	\begin{figure}[tb]
	\includegraphics[]{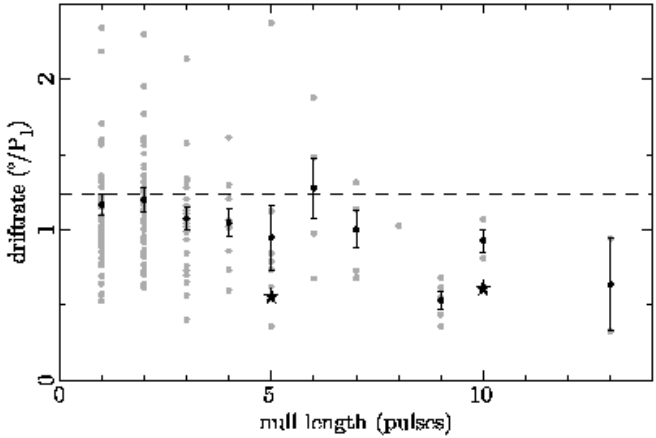}%
	\caption{Drift after the null versus the length of the
	null. We show the average driftrate of the first 6 pulses for
	each driftband after a null (gray points).  The average per
	null length (black points) with its error is also plotted.
	The normal driftrate, corrected for the driftband curvature, is
	indicated by the horizontal dashed line. The stars denote the
	driftrate for the slow drifting mode sequences.  One pulse
	often consists of more than one subpulse, so after many nulls
	we see several driftbands reappear. The number of driftrates
	plotted here is therefore larger than the total number of
	nulls.}
	\label{img:length.vs.slope}
	\end{figure}

If, on the other hand, the subpulse drift would cease abruptly and
completely during the null, the jump in position would be between zero
and two times the average shift in longitude between normal
pulses. The uncertainty in this estimate arises from the fact that
neither the start nor end of the null are known more accurately than to
one pulse period. The mean jump in position over the null would then follow
the horizontal dash-dotted line in Fig. \ref{img:jump.length}.

After discarding the three ambiguous cases (points near the top edge
of Fig. \ref{img:jump.length}), we computed averages for each null
length. These averages follow neither of the two cases outlined
above. There is too much change in position over the null to be
accounted for by just an abrupt stop of the drift, and there is no
evidence for steady (albeit slower than regular) drift during the
null.

We do see that for nulls longer than one, the jump is constant, $1.47
\pm 0.16\:^{o}$ above the offset value that we would expect in the
case of no drift. This independence of null length and subpulse jump
over the null is shown as the dotted horizontal line in Fig.
\ref{img:jump.length}.

\subsubsection{Driftrate around the null}
We have computed the driftband slope around nulls, correcting for
their general curvature.

For the driftrate before the null, we fitted straight lines to the
last six subpulses of each driftband the ended with the null. We find
that this driftrate just before a null does not deviate from the
normal driftrate.

The driftrate after nulls is different from the normal driftrate,
though. Although there is some spread, all the driftrates we find
after longer nulls are lower than the normal average value
(Fig. \ref{img:length.vs.slope}).

\subsubsection{Average pulse profile around null}

	\begin{figure}[b]
	\includegraphics[]{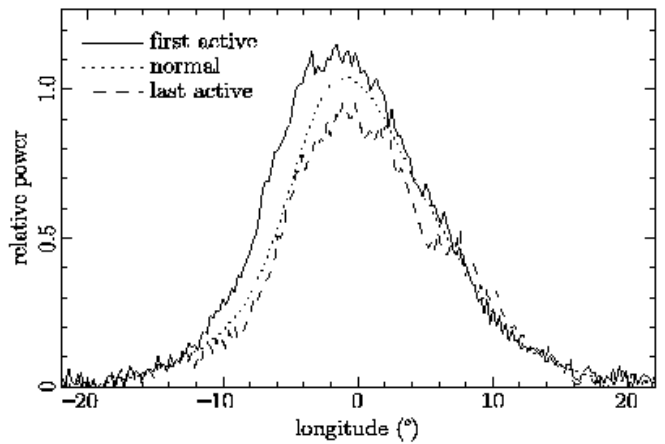}%
	\caption{Average profiles for the pulses adjacent to nulls,
	for all 131 nulls longer than 1 pulse. Shown are: the first
	active pulse after the null, the last active pulse before the
	null and, for reference, the normal average profile.}
	\label{img:int.pro.null}
	\end{figure}

To investigate whether the change from normal emission to the null
state is sudden or more gradual, \citet{la83} compared the energies of
the pulses near a null. The finite chance that the emission from the
pulsars drops or rises within the pulse window would influence the
brightness of the pulses around the null. The last pulse before the
null was indeed found to be less bright than a normal pulse. The first
pulse after the null, however, was considerably more bright than a
normal pulse.

We have not just looked at the energies of these neighbouring pulses,
but also at their profiles. To this end, we have averaged all the last
pulses before nulls longer than 1 period, and all the first pulses
after these nulls. The results, shown in Fig. \ref{img:int.pro.null}
and condensed in Table \ref{tab:ave.prof}, are surprising. Not only do
we find the expected differences in brightness, we also see a
significant offset in the pulse position for the first pulses after
nulls.

\subsection{Slow drifting mode}
\begin{figure}[t]
	\includegraphics[]{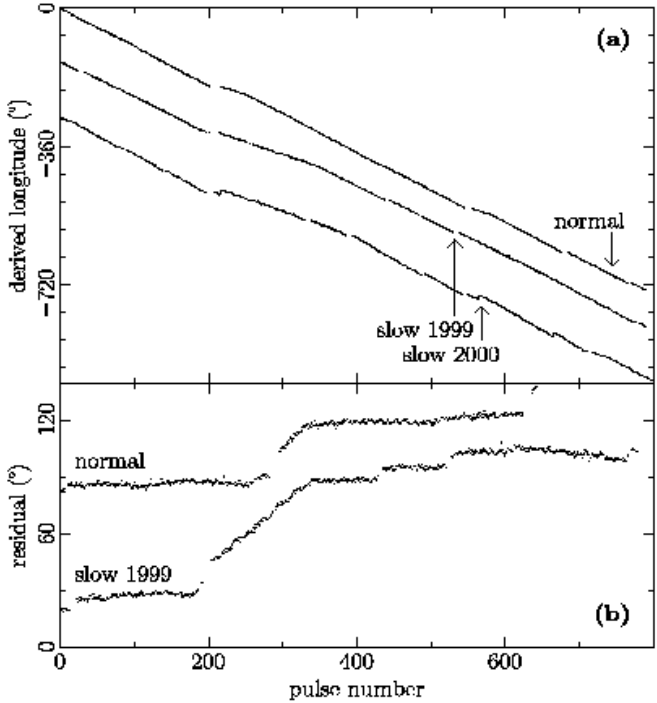}
	\caption{ {\bf a)} Derived longitudes of three pulse
	sequences. The gaps in the curves are nulls. The sequences are
	aligned on the ends of the long nulls around pulse 200.  The
	top line shows the normal drifting behaviour around nulls of
	lengths 13, 9, 3 and 2, respectively.  The middle line shows
	the 1999 slow drifting sequence (offset $-150^{o}$), the
	bottom line shows the 2000 sequence (offset $-300^{o}$). {\bf
	b)} Deviation of the derived longitude from normal drifting
	for the 1999 and the normal sequence shown in panel (a). The
	transitions from slow to normal drifting have been aligned.
	}
	\label{img:der.long}
	\end{figure}

\begin{figure}[tb]
	\centering
	\includegraphics[]{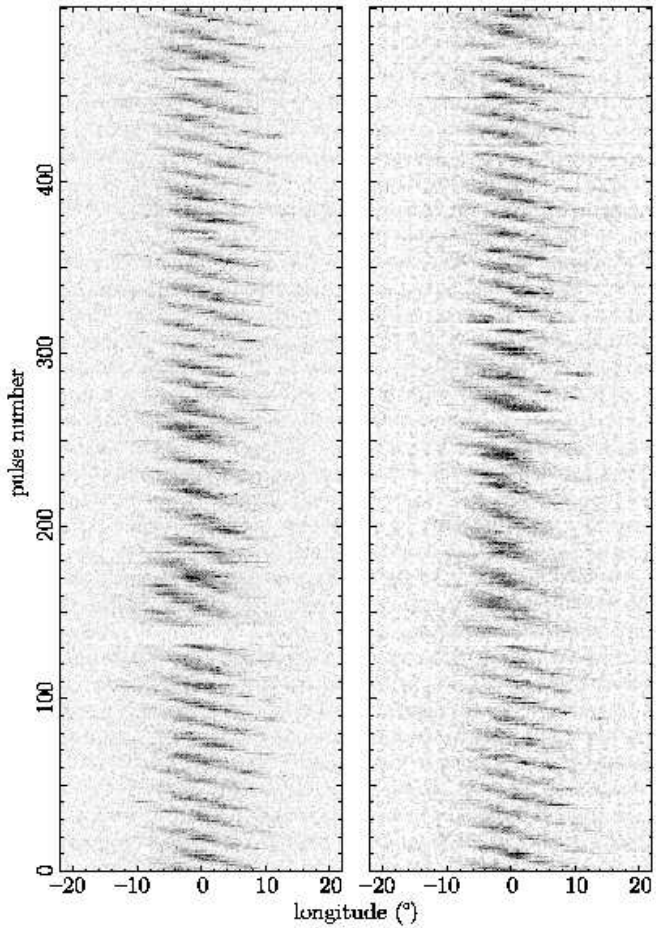}%
	\caption{Grayscale plots of the pulses composing the two slow drifting
	sequences.  The grayscale linearly depicts the intensity
	of a sample. The left series was observed in 1999, the right
	series in 2000. After a null of length 10 (1999 series) and 5
	(2000 series) around pulse number 140, the driftrate and
	driftband separation are steadily different for about 120
	pulses. From pulse 260 on, the drifting changes back to normal.}
	\label{img:slow.obs}
	\end{figure}

Although we had set out to quantify the normally very regular
drifting behaviour of 0809+74, we unexpectedly found two
occasions in which the pulsar clearly deviates from its normal
drifting mode.  We will refer to these sequences by the year in which
they were observed, 1999 and 2000. In Fig. \ref{img:der.long}a we
have plotted the derived longitude \citep{la83} of subsequent pulses
around the mode changes.  The derived longitude of the subpulses
effectively converts the different short driftbands into one long
one. Figure \ref{img:slow.obs} shows a grayscale plot of the two
slow drifting series.

The most striking difference, as the drifting pattern is concerned, is
clearly the decrease of the driftrate by about 50\%.

We see that in both the observations, the drifting mode changes during
or immediately following a long null. After these nulls of length 10
and 5 respectively, almost all drifting properties reappear at new
values, as laid out in Table \ref{tab:mode.prop}. As some of these
properties normally already show change over time, we
compare the slow drifting sequences (labeled `slow') to the 600
pulses that surround them (labeled `normal').

	\begin{table}[b]
	\begin{center}
	\begin{tabular}%
	{|l|r@{\hspace{0.1cm}}c@{\hspace{0.1cm}}l|%
	r@{\hspace{0.1cm}}c@{\hspace{0.1cm}}l|%
	r@{\hspace{0.1cm}}c@{\hspace{0.1cm}}l|}
	\hline
	& \multicolumn{3}{c|}{height}%
	& \multicolumn{3}{c|}{position ($^{o}$)}%
	& \multicolumn{3}{c|}{fwhm ($^{o}$)}\\
	\hline
	\hline
	normal		& 1.000&$\pm$&0.003 &  0.000&$\pm$&0.016& 13.33&$\pm$&0.04\\ 
	last active     & 0.875&$\pm$&0.005 &   0.27&$\pm$&0.12 & 13.96&$\pm$&0.09\\ 
	first active    & 1.112&$\pm$&0.005 &$-$0.77&$\pm$&0.12 & 13.84&$\pm$&0.08\\ 
	\hline
	slow 1999	& 1.272&$\pm$&0.005 &$-$1.51&$\pm$&0.3 &  11.99&$\pm$&0.06 \\ 
	slow 2000	& 1.311&$\pm$&0.008 &$-$0.75&$\pm$&0.3 &  10.52&$\pm$&0.08 \\ 
	\hline
	\end{tabular}
	\caption{Comparison of the pulsar's average emission profiles
	for different sets of pulses. We show the height, full width
	at half maximum (fwhm) and position of the Gaussians that
	fitted the profiles best. The `normal' subset consist of all
	non-null pulses in the dataset. All heights and positions in
	this table are relative to the height and position of this
	`normal' set. The set labeled `last active' contains
	all 131 pulses that preceded a null that was longer than 1
	pulse period. The characteristics of the pulses that followed
	these nulls are labeled `first active'. The subsets
	`slow 1999' and `slow 2000' contain the 120 pulses that
	form the slow drifting sequences observed in 1999 and 2000,
	respectively.}
	\label{tab:ave.prof} 
	\end{center} 
	\end{table}

In the first three columns of Table \ref{tab:mode.prop}, we
investigate the parameters of the individual subpulses in both
modes. Immediately we see a very interesting change in the relative
average position of the subpulses. Right after the null, there is a
definite shift in position towards earlier arrival, that can already
be seen in a plot like Fig. \ref{img:slow.obs}.  Interestingly, this
shift of the pulse window is similar to the average jump in subpulse
position we see over a normal null.

In the slow drifting mode, the subpulses are slightly wider, but their
heights remain the same.

Next, we explore the driftband characteristics in the last three
columns of Table \ref{tab:mode.prop}. The average longitude
separation of two adjacent driftbands, $P_2$, decreases significantly
by about 15\%. Hence, in the slow drifting mode the subpulses within
each pulse are spaced closer together than normal.

When we compare the driftrates of the subpulses between the two modes,
we find that the driftrate in the slow drifting mode is almost
halved. This puts the new driftrate right in the range of driftrates
we usually find after a longer null. We have indicated these driftrate
values with black stars in Fig.  \ref{img:length.vs.slope}.

Being dependent on $P_2$ and the driftrate, the fractional change in
$P_3$ (the recurrence time of a driftband) is comparably large.

In both cases the new drifting mode is stable for about 120 pulses and
then changes back to normal. To illustrate this change, we have
plotted the deviation of the derived longitude from normal drifting in
Fig. \ref{img:der.long}b. For the 1999 observation, we see a normal
driftrate up to the null at pulse 200. After this null, the driftrate
is smaller, up to pulse 330. In about 20 pulses, the drift
then speeds up back to its normal value.

For reference, we have also plotted the drifting behaviour after the
longest null in our sample. Again, we see a speedup to the normal
driftrate in about 20 pulses at pulse 330, very similar to the slow
drifting speedup time scale.

The slow drifting interval in the 2000 observation (right panel of
Fig. \ref{img:slow.obs}) is followed and stopped by a series of
frequent, longer than average nulls, out of which the pulsar emerges
in its normal drifting pattern.

These surprising changes in subpulse positions, widths and separations
must influence the average pulse profile during the slow drifting
mode. Noting the many similarities between slow drifting pulses and
the pulses that follow nulls (halved driftrate, similar offsets,
similar speedup) we are interested in the possible similarities
between the average profiles of these pulses, especially since the
average profile of the pulses after the null is singularly bright and
shifted in longitude.

\begin{figure}[tb]
	\includegraphics[]{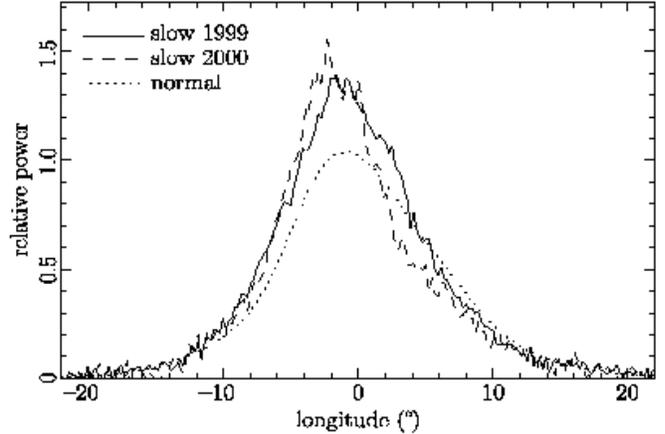}
	\caption{Average profiles for the slow drifting mode
	sequences. The `slow 1999' and `slow 2000' profiles are
	averaged over the 120 pulses that composed the slow drifting
	sequences observed in 1999 and 2000, respectively. The normal
	average profile is plotted for comparison.}
	\label{img:int.pro.slow}
	\end{figure}

In Fig. \ref{img:int.pro.slow} and Table \ref{tab:ave.prof} we compare
the pulsar's normal average profile with the average profile in the
slow drifting mode. In both observations, the slow drifting mode
average profile is much brighter than the normal profile and offset
towards earlier arrival -- exactly the two peculiarities we found in
the average pulse profile after a null as well.

\section{Discussion}
\subsection{Driftband fitting}
The non-linearity of the driftbands was already noted by
\citet{pag73}, who integrated several hundreds of pulses to compare
the shape of the driftbands in different observations. He found
considerable differences between these driftband shapes.  We compare
the average drift path over several thousands of pulses however, and
find that the curvature of the driftbands remains the same.

The curvature of the driftband can be a direct consequence of the
curved geometry of the emission region \citep{kri80}. In that case,
the shape of the driftband is expected to be point-symmetric: the
curve will then be unchanged after mirroring it in both axes. This
point-symmetry would then return in the residuals to straight line
fits (Fig. \ref{img:resid}). The shape of the driftbands we find,
however, is not point-symmetric but axisymmetric: the curve is
unchanged after mirroring in the y-axis.

We could still tie the curvature of the driftbands to the
emission-region geometry, by assuming that part of the pulse profile
of this pulsar is missing. In that case, the driftband shape we find
would represent only part of the total expected shape.

This suggestion that part of the profile of 0809+74 is missing
fits in with observations of this pulsar's pulse shape at different
frequencies \citep{bkk+81,kis+98}. These imply that at lower
frequencies the profile is partially ``absorbed''.

\subsection{Shortest nulls}
We find that the number of invisible nulls is low. That means we can
study of the drifting properties of 0809+74 straightforwardly.

	\begin{table*}
	\begin{center}
	\begin{tabular}%
	{|l|
	r@{\hspace{0.1cm}}c@{\hspace{0.1cm}}l|
	r@{\hspace{0.1cm}}c@{\hspace{0.1cm}}l|
	r@{\hspace{0.1cm}}c@{\hspace{0.1cm}}l|
	r@{\hspace{0.1cm}}c@{\hspace{0.1cm}}l|
	r@{\hspace{0.1cm}}c@{\hspace{0.1cm}}l|
	r@{\hspace{0.1cm}}c@{\hspace{0.1cm}}l|}
	\hline
	& \multicolumn{9}{c|}{averages from Gaussian fits to subpulses}
	& \multicolumn{9}{c|}{averages from straight line fits to driftbands}
	\\
	\cline{2-19}
	& \multicolumn{3}{c|}{relative height} 
	& \multicolumn{3}{c|}{relative position ($^{o}$)} 
	& \multicolumn{3}{c|}{fwhm ($^{o}$)}
	& \multicolumn{3}{c|}{$P_2$ ($^{o}$)}
	& \multicolumn{3}{c|}{driftrate ($^{o}$/$P_1$)}
	& \multicolumn{3}{c|}{$P_3$ ($P_1$)}
	\\
	\hline
	\hline
	normal 1999
	       	& ~~ 1.00  &$\pm$& 0.02 ~~      
	      	& ~~~~~ 0.00  &$\pm$& 0.11 ~~~~~
	      	& 5.39  &$\pm$& 0.06		
	      	& 11.61  &$\pm$& 0.14		
		& ~~1.086  &$\pm$& 0.011~~	
		& 10.72 &$\pm$& 0.13\\		
	slow 1999
	        & 1.06  &$\pm$& 0.05   		
		& $-1.2$&$\pm$& 0.3  		
		& 5.75  &$\pm$& 0.11  		
	        & 9.9   &$\pm$& 0.5 		
	        & 0.606 &$\pm$& 0.019  	        
		& 16.5  &$\pm$& 0.5\\   	
	\hline
	\hline
	normal 2000
		& 1.00  &$\pm$& 0.02   		
		& 0.0   &$\pm$& 0.2   		
		& 5.22  &$\pm$& 0.06   		
		& 11.33 &$\pm$& 0.14   		
		& 1.043 &$\pm$& 0.014    	
		& 10.88 &$\pm$& 0.13\\		
	slow 2000
		& 0.99  &$\pm$& 0.04   		
		& ~$-1.8$ &$\pm$& 0.2 ~   	
		& 5.55  &$\pm$& 0.11   		
		& 9.63  &$\pm$& 0.3		
		& 0.54  &$\pm$& 0.02		
		& 17.6  &$\pm$& 0.5\\ 		

	\hline
	\end{tabular}
	\caption{Subpulse and driftband properties for the
 	slow drifting mode sequences and the surrounding normal
 	drifting pattern.}
	\label{tab:mode.prop}
	\end{center} 
	\end{table*}

\subsection{Slow drifting mode and nulling}
With our investigation of 0809+74 still ongoing, we discuss here only
the phenomena themselves and defer their interpretation to a
subsequent paper.

The similarity between this pulsar's behaviour in the slow drifting
mode and around nulls is striking:
\begin{itemize}
\item	The slow drifting sequences start at nulls.
\item	The driftrate of the slow drifting mode is a lower limit
	to the driftrates found after all nulls.
\item 	The speedup from slow drifting to normal is
	identical to the speedup after a null.
\item	The average profiles of both the post-null and slow drifting
	pulses are brighter than normal.
\item	These average profiles are both displaced to earlier arrival.
\item	This displacement of the  average profile is caused by offsets in the
	individual subpulses.
\item	These offsets are identical to the jump of the subpulses over the
	null.
\end{itemize}

The natural conclusion is that the behaviour after a null and the slow
drifting mode are the same, quasi-stable phenomenon. Normally the
pulsar reappears from the null in the slow drifting mode. After a
variable time, it quickly evolves to the normal mode. Therefore, right
after a normal null we see either a short sequence of slow drifting,
or the transition back to normal drifting. In the case of the long
slow drifting mode sequences, the metamorphosis back to the normal
mode is delayed.

This would explain all the similarities we found above. After a normal
null, the pulsar is in the slow drifting mode or changing back to the
normal mode. All the characteristics of the slow drifting mode can
then be identified in the post-null behaviour, although they will be
less pronounced; the transition to normal drifting may already be
taking place.

When the driftrate increases at the end of a slow drifting sequence,
this speedup is quick and identical to the speedup seen after a normal
null (see Fig. \ref{img:der.long}b).

The post-null driftrate values are found in between the slow drifting
and normal driftrate value, depending on how soon the transition back
to normal takes place (see Fig. \ref{img:length.vs.slope}). Although
the return from the low to the normal driftrate will be quick, the
time the driftrate is low may vary for different nulls of the same
length. The average of these different sequences  will
then resemble the slow exponential decay found by \citet{la83}.

Comparing average pulse profiles, the increased brightness and the
pulse offset of the post-null average profile are attenuated versions
of similar deviations seen in the slow drifting mode profile. (see
Table \ref{tab:ave.prof}, Figs. \ref{img:int.pro.null} and
\ref{img:int.pro.slow})

The average profile offset we see in the slow drifting mode is caused
by a shift of the window in which the subpulses appear (Tables
 \ref{tab:mode.prop} and \ref{tab:ave.prof}).

The change in position of the post-null average pulse profile must
then be caused by this shift of the pulse window as well. The
magnitude of this shift is identical to the subpulse-longitude jump
over normal nulls.

This means that the subpulse-position jump over a null is caused by a
displacement of the pulse window as a whole, like in the slow drifting
mode.

Previously, this jump was thought to be the effect of the
subpulse-drift speedup during the null. For this speedup to produce
the observed jump in subpulse position, the time scale involved
had to be long, contrasting the short time scales found for the
slowdown of the subpulse drift and the rise and fall of the emission
around a null.

With the displacement of the subpulses over the null accounted for,
the estimated speedup time of the subpulse drifting is negligible: the
preservation of the position of the subpulses over the null now only
allows for a quick speedup of the subpulse drift, putting all
time scales of emission and drift rise and decay around the nulls in
the same range.

\section{Conclusions}
After many or all nulls, 0809+74 emits in a mode different from
the normal one. This mode is quasi-stable, normally changing back to
normal in several pulses. This is seen as the normal behaviour after a
null, of which the reduced driftrate is the most striking
characteristic. Occasionally, the quasi-stable slow drifting
configuration persists for over a hundred pulses before
changing back to the normal mode.

The pulses in the slow drifting mode and consequently all pulses
after the null are brighter than normal pulses. In the slow drifting
mode, the subpulses are closer together, they drift more slowly
through the profile and the window in which they appear is offset
towards earlier arrival.

This offset of the pulse window accounts for the displacement of
subpulses over the null. When taking this shift of the window into
account, we find that the longitude of the subpulses is perfectly
conserved over a null. This indicates that the speedup time for the
subpulse drift is short.


\begin{thebibliography}{24}
\expandafter\ifx\csname natexlab\endcsname\relax\def\natexlab#1{#1}\fi

\bibitem[{Alexeev {et~al.}(1969)Alexeev, Vitkevich, \& Shitov}]{avs69}
Alexeev, Y., Vitkevich, V., \& Shitov, Y. 1969, Astron. Circ. Acad. of Sci.
  USSR, N495, 4

\bibitem[{Bartel {et~al.}(1981)Bartel, Kardashev, Kuzmin, Nikolaev, Popov,
  Sieber, Smirnova, Soglasnov, \& Wielebinski}]{bkk+81}
Bartel, N., Kardashev, N.~S., Kuzmin, A.~D., {et~al.} 1981, A\&A, 93, 85

\bibitem[{Cole(1970)}]{col70a}
Cole, T.~W. 1970, Nature, 227, 788

\bibitem[{Davies {et~al.}(1984)Davies, Lyne, G., Izvekova, Kuzmin, \&
  Shitov}]{dls+84}
Davies, J.~G., Lyne, A.~G., G., S.~F., {et~al.} 1984, MNRAS, 211, 57

\bibitem[{Deshpande \& Rankin(1999)}]{dr99}
Deshpande, A.~A. \& Rankin, J.~M. 1999, ApJ, 524, 1008

\bibitem[{Deshpande \& Rankin(2001)}]{dr01}
---. 2001, MNRAS, 322, 438

\bibitem[{Drake \& Craft(1968)}]{dc68}
Drake, F.~D. \& Craft, H.~D. 1968, Nature, 220, 231

\bibitem[{Filippenko \& Radhakrishnan(1982)}]{fr82}
Filippenko, A.~V. \& Radhakrishnan, V. 1982, ApJ, 263, 828

\bibitem[{Goldreich \& Julian(1969)}]{gj69}
Goldreich, P. \& Julian, W.~H. 1969, ApJ, 157, 869

\bibitem[{Kouwenhoven(2000)}]{kou00}
Kouwenhoven, M. L.~A. 2000, PhD thesis, Utrecht University

\bibitem[{Krishnamohan(1980)}]{kri80}
Krishnamohan, S. 1980, MNRAS, 191, 237

\bibitem[{Kuzmin {et~al.}(1998)Kuzmin, Izvekova, Shitov, Sieber, Jessner,
  Wielebinski, Lyne, \& Smith}]{kis+98}
Kuzmin, A.~D., Izvekova, V.~A., Shitov, Y.~P., {et~al.} 1998, A\&AS, 127, 255

\bibitem[{Lyne \& Ashworth(1983)}]{la83}
Lyne, A.~G. \& Ashworth, M. 1983, MNRAS, 204, 519

\bibitem[{Page(1973)}]{pag73}
Page, C.~G. 1973, MNRAS, 163, 29

\bibitem[{Popov \& Smirnova(1982)}]{ps82}
Popov, M.~V. \& Smirnova, T.~V. 1982, Sov. Astron., 26, 439

\bibitem[{Press {et~al.}(1992)Press, Teukolsky, Vetterling, \&
  Flannery}]{ptvf92}
Press, W.~H., Teukolsky, S.~A., Vetterling, W.~T., \& Flannery, B.~P. 1992,
  Numerical Recipes: {T}he Art of Scientific Computing, 2$^{nd}$ edition
  (Cambridge: Cambridge University Press)

\bibitem[{Ritchings(1976)}]{rit76}
Ritchings, R.~T. 1976, MNRAS, 176, 249

\bibitem[{Ruderman \& Sutherland(1975)}]{rs75}
Ruderman, M.~A. \& Sutherland, P.~G. 1975, ApJ, 196, 51

\bibitem[{Sturrock(1971)}]{stu71}
Sturrock, P.~A. 1971, ApJ, 164, 529

\bibitem[{Taylor \& Huguenin(1971)}]{th71}
Taylor, J.~H. \& Huguenin, G.~R. 1971, ApJ, 167, 273

\bibitem[{Unwin {et~al.}(1978)Unwin, Readhead, Wilkinson, \& Ewing}]{urwe78}
Unwin, S.~C., Readhead, A. C.~S., Wilkinson, P.~N., \& Ewing, M.~S. 1978,
  MNRAS, 182, 711

\bibitem[{Vitkevich \& Shitov(1970)}]{vs70}
Vitkevich, V. \& Shitov, Y. 1970, Nature, 225, 248

\bibitem[{Vo\^{u}te(2001)}]{vou01}
Vo\^{u}te, J. L.~L. 2001, PhD thesis, University of Amsterdam

\bibitem[{Vo\^{u}te {et~al.}(2002)Vo\^{u}te, Kouwenhoven, van Haren, Langerak,
  Stappers, Driesens, Ramachandran, \& Beijaard}]{vkh+02}
Vo\^{u}te, J. L.~L., Kouwenhoven, M. L.~A., van Haren, P.~C., {et~al.} 2002,
  A\&A, accepted

\end{thebibliography}

\end{document}